# Three Dimensional Measurements by Deflectometry and Double Hilbert Transform

Silin Na*, Sanghoon Shin**, Younghun Yu*

* Department of Physics, Jeju National University, Jeju, 63243, Korea

** Kanghae Precision System, Hawsung, 18487, Korea

An improved phase retrieval method based Hilbert transform is introduced to quantitatively calculate the phase distribution from distorted fringe pattern. Also phase measurement deflectomety are widely used in specular type samples. The background noise or bias should be suppressed prior to apply Hilbert transform. A method for suppression background noise double Hilbert transform is presented, which requires only one image. The method is easy to implement, and it is able to conducting automated fast measurements. We have demonstrated the double Hilbert transform method to retrieve the phase and background suppression by computer simulation and experiment in phase measuring deflectometry method.

PACS Numbers.; 42. 30.-d, 42. 30. Rx, 42. 30. Kq

Keywords: Deflrctometry, Phase measuring deflectometry, Hilbert transform

I.   Introduction

Digital fringe method to retrieve the surface topography of three-dimensional (3-D) object is one of the active research areas in optical metrology [1-3]. In these methods, fringe patterns are obtained as the output of a measuring system. The spatially varying phase of fringes patterns are related to physical quantity of samples. Its application ranges from measuring the 3-D surface of optical components to measuring the industrial mechanical system. It is whole field, non-contact, non-invasive, inexpensive, and providing high resolution. In recent years, digital image processing and pattern recognition techniques are widely used for automatic fringe analysis. Phase shifting is traditional method for phase retrieve from deformed fringes [4-6]. This method requires at least three images for retrieving phase. So it is not adaptable for fast measurements. For fast measurement, single frame methods are necessary. Single frame methods such as Fast Fourier Transform (FFT) and Hilbert transform (HT) have been suggested [7-10]. In FFT method the image is processed in the frequency domain, but HT method can be implemented in spatial domain. FFT methods, since Takeda proposed, carried out to improve the performances and quality. FFT method works well for open fringe patterns. However, this method has a major drawback: it is necessary manually identify background noise and signal in frequency space.

HT methods has not been much worked on digital fringe method. Compare to FFT, HT method has many advantages, simplicity and can be fully automatic. The success of use HT of the signal for retrieving phase depends on preservation of signal amplitude and elimination background noise. There are large number of methods for background elimination and preserving the constant amplitude in captured image [11-14]. However, the methods need large calculation time and complex, and could not use in fast and automatic system.

In this paper we have used double HT method to eliminate background and experimentally demonstrated using phase measurment deflectometry.

## II. Theoretical Model

II-1. Deflectometry

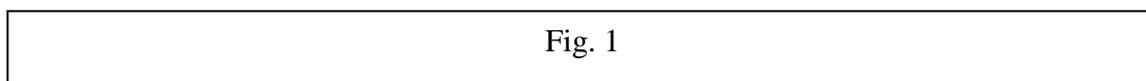

Fig. 1

Deflectometry is useful measurement method for specular type samples. Consider the optical arrangement shown in Fig. 1. The CCD camera is focused on the screen on which a periodic fringe pattern is displayed. The irradiance distribution captured by the camera, $I(x,y)$, is given by

$$I(x, y) = I_0(x, y) + b(x, y) \cos\left(\frac{2\pi(x+y)}{p} + \phi(x, y)\right) \qquad (1)$$

where $p$ is the period of fringe pattern and $\phi(x, y)$ is the phase, which relate the slope of sample. When the phase object is positioned on the optical path, it changes the optical path length. If the phase is inhomogeneous in the x-direction, the ray will be deflected by an angle $\Phi(x,y) \approx \partial W(x,y)/\partial x$, where $W(x,y)$ is the optical path length of a ray traveling through the phase object at the position $(x,y)$ [17]. The phase information gives the gradient of the optical path. We can calculate the shape of the sample by integration or least square method using gradient data [15-17].

II-2. Double Hilbert transform for elimination background noise

The background intensity $I_0(x, y)$ which is uniform or non-uniform, must be eliminated prior to the application HT for retrieving phase. There are many researches to suppress the bias. However, the suppression methods are complex and time consumption process.

Hilbert transform is a linear operator which takes a function f(t), and produce a function, HT(f(t)), in the same domain;

$$HT(f(t)) = \frac{1}{\pi} \int \frac{f(t)}{(t-\tau)} d\tau \qquad (2)$$

Hilbert transform is equivalent to their phase shift by $\frac{\pi}{2}$ and filtering DC component without altering the amplitude [18,19]. The HT of the signal $f(x) = \cos(x)$ gives $-\sin(x)$. In Eq. (1), background noise, $I_0(x, y)$, should be suppressed to apply HT method. The HT of Eq. (1) can be written as follow;

$$HT(I(x,y)) = b(x,y)\sin\left(\frac{2\pi(x+y)}{p} + \phi(x,y)\right) \qquad (3)$$

We can show that the bias was eliminated by HT. Doing HT again on the result as shown in Eq. (3), we can obtain

$$HT(HT(I(x,y))) = -b(x,y)\cos\left(\frac{2\pi(x+y)}{p} + \phi(x,y)\right) \qquad (4)$$

From Eq. (3) and Eq. (4), the phase distribution of the fringe pattern can be solved by

$$\phi'(x,y) = -\arctan\frac{HT(I(x,y))}{HT(HT(I(x,y)))} \qquad (5)$$

It is easier and simple method for retrieving the phase information from distorted image.

The calculated value $\phi'(x)$ in Eq. (5) differs from real value $\phi(x)$ due to finite number of fringe cycle. This error can be corrected as

$$\phi(x) = \phi'(x) + \varepsilon \qquad (6)$$

where $\varepsilon$ is the error. The error $\varepsilon$ can be calculated for any value $\phi$ [8,10].

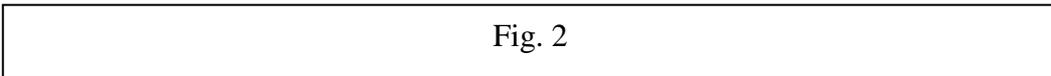

Fig. 2

Figure 2 shows the error between input phase and calculated phase using hilbert transform. This error is used to correct the calculated phase, $\phi'$.

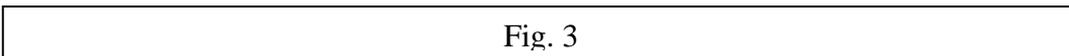

Fig. 3

Figure 3 shows the computer simulation results for lens using Hilbert transform. Figure 3(a) and (b) show the original distorted pattern image and 2-dimensional Fourier transformed image. From Fig. 3 (b), we can show the bias ($0^{th}$) is exists. Figure 3(c) and (d) show the

Hilbert transformed image of Fig. 3(a) and 2-dimensional Fourier transformed image. Compare Fig. 3(b) and (d), we can show that the bias was eliminated by Hilbert transform. Figure 3(e) shows the double Hilbert transformed image and Fig. 3(f) shows the profile of dotted line in Fig. 3(c) and (d). The line is the profile of dostted line in Fig. 3(c) and dotted line os that of Fig. (e). From Fig. 3(f), the phase is shifted $\frac{\pi}{2}$ by Hilbert transform.

### III. Experiment and Results

Fig. 4

Figure 4 shows the schematic transmission type phase measuring deflectometry experimental set-up.

We used a CCD camera (Imperx) to record the distorted images. The pixel size and the number of pixels were 7.4 × 7.4 μm and 1024 × 1024, respectively. And we used an LCD monitor for display, a conic lens was used as phase object. The distance between the lens and display was 45 mm, and the distance between the screen and camera was 650 mm. The period of the fringe pattern was 2.5 mm

Fig. 5

Figure 5 shows the experimental results for conic lens. Figure 5 (a), (b) and (c) are captured distorted pattern image, profile of dotted line in (a) and Fourier transformed image. We can show that the bias term is exists from Fig. 5 (c). Figure 5 (d), (e) and (f) are Hilbert transformed image of Fig. 5 (a), double Hilbert transformed image and profile of dotted line in Fig. 5 (c) and (d). From Fig. 5 (f), we can show that the bias is suppressed and phase is shifted by $\frac{\pi}{2}$. We have done same analysis with y-axis distorted image. The phase can be calculated by using HT image and double HT image. The calculation result is shown in Fig.6.

Figure 6 shows the calculated phase and height using Fig. 5. Figure 6 (a) and (b) are x-axis and y-axis distorted pattern images, and Fig. 6(c) is the 3D gray level height image, and Fig. 6(d) is the profiles of dotted line in Fig. 6(c). We obtained the unwrapped phases following the usual unwrapping process, and use least square method for height calculation [17,20]. The measured slope of conic lens is $4.98 \pm 0.03$ degree, which value is very similar to design value, 5 degree.

Fig. 6

## IV. Conclusion

Digital phase measuring deflectometry are widely used in three dimensional shape measurements of specular materials. To retrieve the phase information which is related height of sample, FFT and HT methods are used. These methods need only one fringe image and is used in fast measurement. The background intensity, which is uniform or non-uniform, must be eliminated prior to the application HT for retrieving phase. We used double HT method to eliminate background noise. It is easy and simple method for elimination background noise and retrieving the phase. We have verified that the background noise is eliminated by double HT using computer simulation. Also we have demonstrated this method by transmission type deflectometry experiments and verified that this method is useful for three dimensional shape measurements.

**Acknowledgment**

This research was supported by the 2016 scientific promotion program funded by Jeju National University

Figure Captions

Fig. 1. Schematic diagram of phase measurement deflectometry.

Fig. 2. Calibration curve. (a) Calibration curve between the input phase and calculated phase, and (b) phase error

Fig. 3. Computer simualtion result. (a) original distorted pattern image, (b) 2-dimensional Fourier transformed image of (a), (c) Hilbert transformed image, (d) 2-dimensional Fourier transformed image of (c), (e) double Hilbert transformed image, (f) profiles of HT and double HT image

Fig. 4. Schematic diagram of the experimental setup for transmission-type deflectometry, using the spatial light modulator (SLM).

Fig. 5. Experimental results of conic lens. (a) distorted X-axis image, (b) profile of dotted line in (a), (c) 2-domensional Fourier transformed image if (a), (d) Hilbert transformed image, (e) double Hilbert transformed image of (a) , (f) profiles of HT and double HT image

Fig. 6. Captured pattern image and calculated height. (a)(b) x-axis and y-axis deformed image, (c ) calculated 3D grey level image, (d) profile dotted line in (c)

Figures

Fig. 1

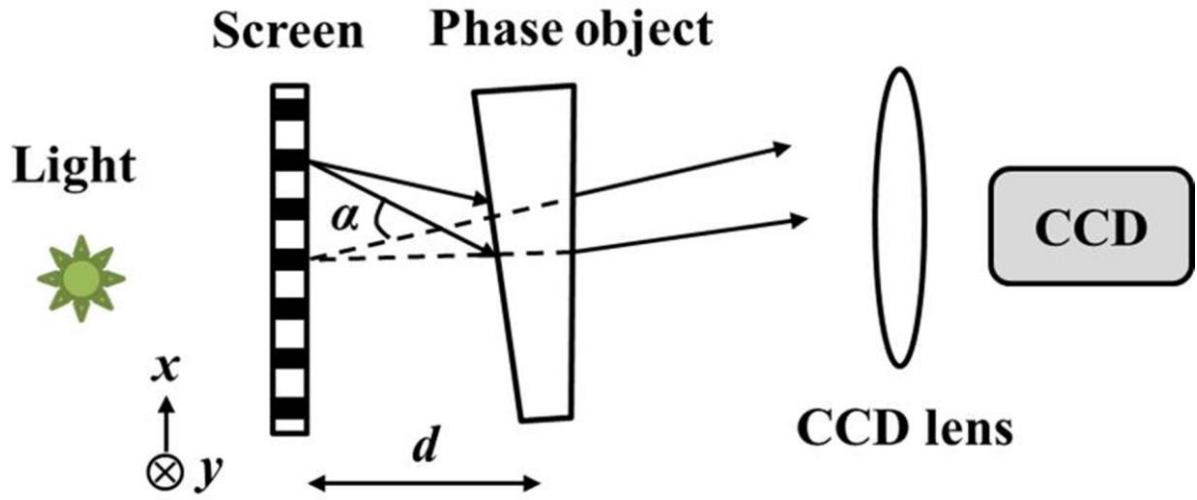

Fig. 2

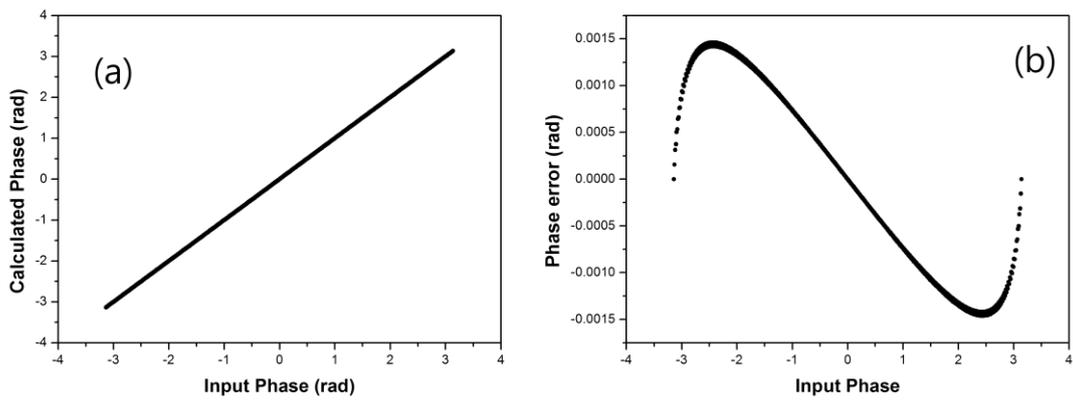

Fig. 3

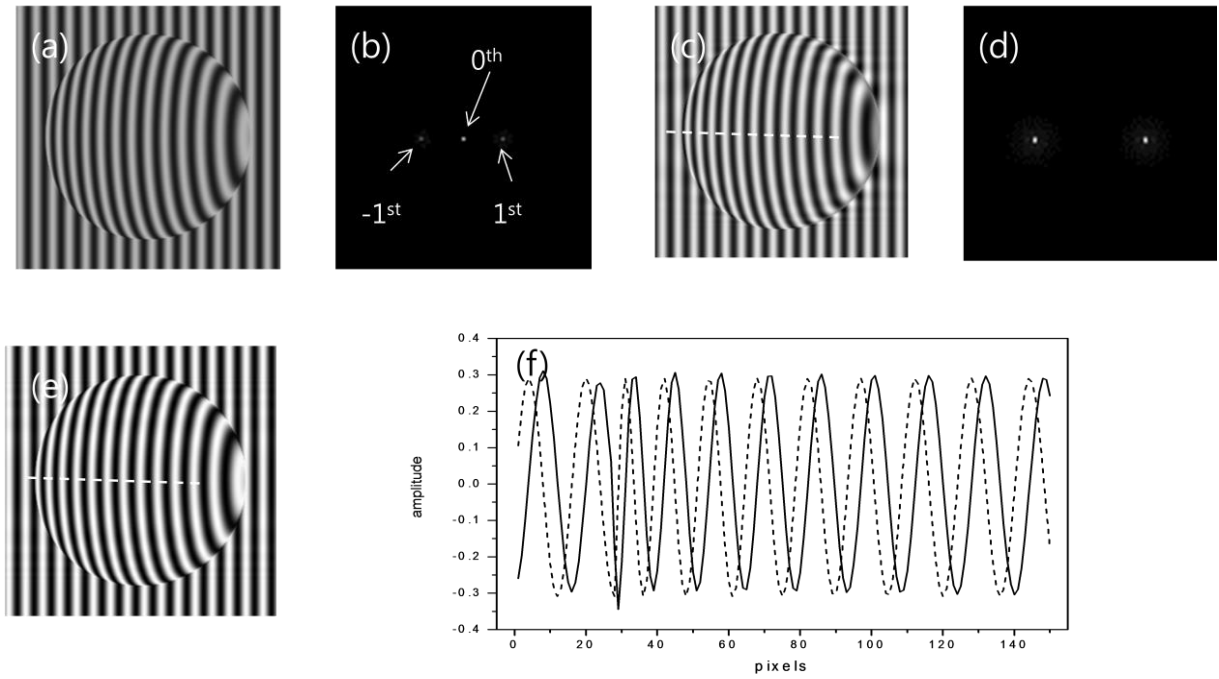

Fig. 4

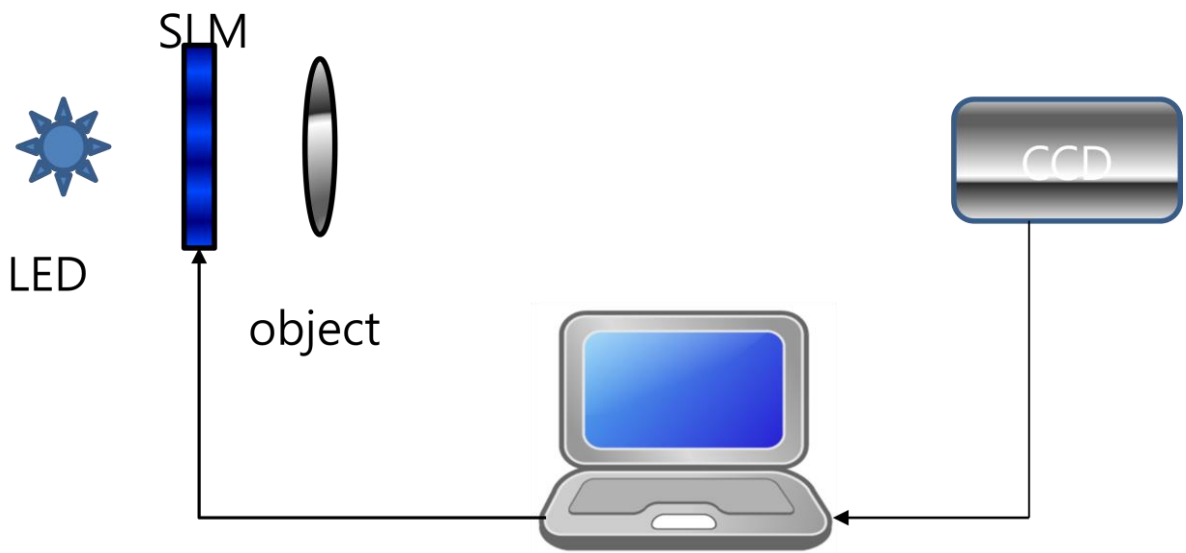

Fig. 5

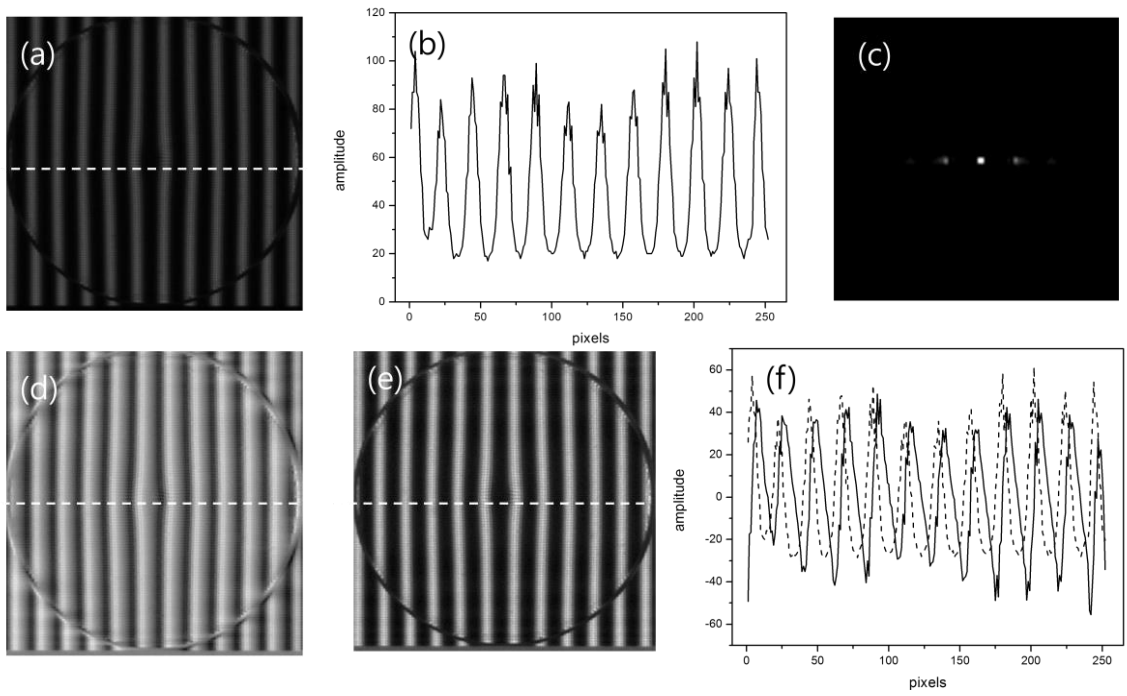

Fig. 6

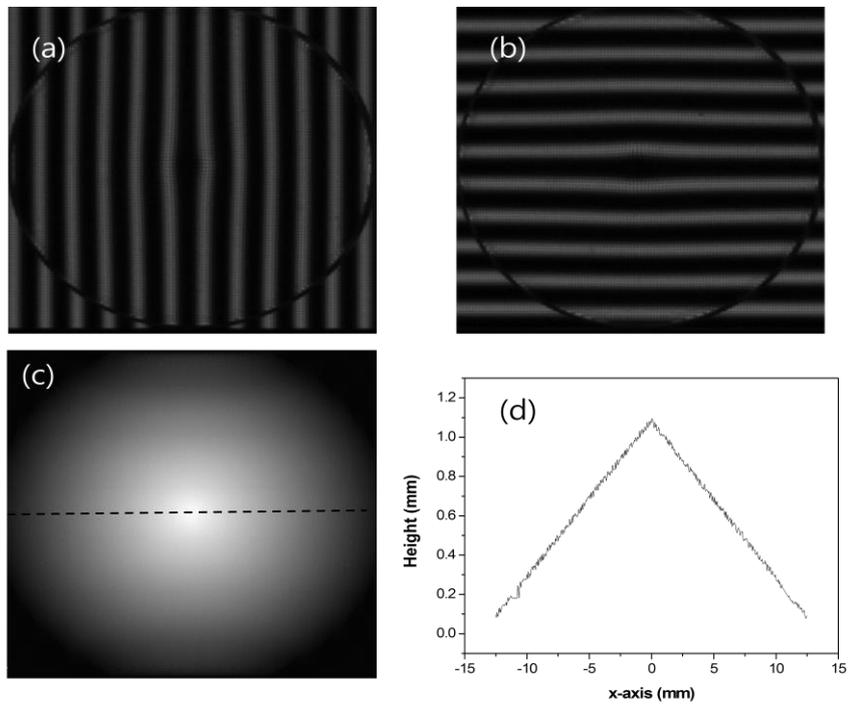